\documentstyle[psfig,aps,floats,eqsecnum]{revtex}
  \newlength\titlebox 
\twocolumn 
\narrowtext
\newcommand{\be}{\begin{eqnarray}}
\newcommand{\ee}{\end{eqnarray}}

\newcommand\ie {{\it i.e.\ }}
\newcommand\eg {{\it e.g.\  }}

\newcommand\cf {{\it cf.\   }}

\newcommand{\del}{\partial}

\newcommand{\la}{\langle}
\newcommand{\ra}{\rangle}
\begin{document}
\preprint{KRL MAP-197}
\vskip 1cm
\title{Propagation of Information in Populations of Self-Replicating Code}
\author{Johan Chu$^{(1)}$ and Chris Adami$^{(2,3)}$}
\address{\mbox{}\\
$^{(1)}$Department of Physics 114-36 \\
$^{(2)}$W. K. Kellogg Radiation Laboratory 106-38 \\
$^{(3)}$Computation and Neural Systems 139-74	\\
California Institute of Technology, Pasadena, CA 91125 }
\maketitle
\begin{abstract}
We observe the propagation of information in a system of 
self-replicating strings of code (``Artificial Life'') as a function
of fitness and mutation rate. Comparison with theoretical predictions
based on the reaction-diffusion equation shows that the response of the
artificial system to fluctuations (\eg velocity of the information wave 
as a function of relative fitness) closely follows that of natural systems. 
We find that the 
relaxation time of the system depends on the speed of propagation of
information and the size of the system. This analysis offers the possibility 
of determining the minimal system size for observation of non-equilibrium 
effects at fixed mutation rate. 
\end{abstract}
\narrowtext
\section{Introduction}
Thermodynamic equilibrium systems respond to perturbations with waves that 
re-establish
equilibrium. This is a general feature of statistical systems, but it can
also be observed in natural populations, where the disturbance of interest 
is a new 
species with either negligible or positive fitness advantage. The new species
spreads through the population at a rate dependent on its relative fitness  
and some basic properties of the medium which can be summarized by the 
diffusion coefficient. This problem has been addressed theoretically~\cite{FIS}
and experimentally (see \eg \cite{DOB} and references therein) since 
early this century. The application of the appropriate machinery (diffusion 
equations) to the spatial propagation of {\em information} rather than 
species, is 
much more recent, and has been successful in the description of 
experiments with {\em in vitro} evolving RNA~\cite{BMO,MB}. 

Systems of self-replicating information (\cf the replicating
RNA system mentioned above) 
are often thought to represent the simplest living system. They offer the 
chance to isolate the mechanisms involved in information transfer (from
environment into the genome) and propagation (throughout the population),
and study them in detail.

It has long been suspected that living systems operate, in a thermodynamical 
sense, far away from the equilibrium state. On the molecular scale, many 
of the chemical reactions occurring in a cell's metabolism require 
non-equilibrium conditions. On a larger scale, it appears that only 
a system far away from equilibrium can produce the required diversity 
(in genome) for evolution to proceed effectively (we will comment on this 
below).

In the systems that we are interested in -- systems of self-replicating 
information in a noisy and information-rich environment --
the processes that work for and against equilibration of information
are clearly mutation and replication. In the 
absence of mutation, replication leads to a uniform non-evolving state 
where every member
of the population is identical. Mutation in the
absence of replication, on the other hand, leads to maximal diversity of the 
population but no evolution either, as selection is absent. Thus, effective
adaptation and evolution depend on a balance of these driving 
forces
(see, \eg \cite{CA1,CA3}). The relaxation time of such a system, however,
just as in thermodynamical systems, is mainly dictated by the mutation rate
which plays the role of ``temperature'' in these systems~\cite{CA3}. As
such, it represents a crucial parameter which determines how close the 
system is to ``thermodynamical'' equilibrium. Clearly, a relaxation time
larger than the average time between (advantageous) mutations will result 
in a non-equilibrium system, while a smaller relaxation time leads to fast
equlibration.  The relaxation time may be defined as the time it takes 
information to spread throughout the entire system ({\it i.e.} travel an 
average
distance of half the ``diameter'' of the population). A non-equilibrium
population therefore can always be obtained (at fixed mutation rate) 
by increasing the size of the system. At the same time, such a large system
segments into areas that effectively cannot communicate with each other, but 
are close to equilibrium themselves. This may be the key to genomic diversity,
and possibly to speciation in the absence of niches and explicit barriers.

The advent of artificial living systems such as {\sf tierra}~\cite{RAY1,CA1}
and 
{\sf avida}~\cite{AB1,ABH} have opened up the possibility of checking these 
ideas 
explicitly, as the evolutionary pace in systems both close and far away from
equilibrium can be investigated directly. As a foundation for such 
experiments, in this paper we investigate  the dynamics of information 
propagation in the artificial life system {\sf sanda}, a variant of the
{\sf avida} system designed to run on arbitrarily many parallel processors.
This is a necessary capability for investigating arbitrarily large 
populations of 
strings of code. The purpose of our experiments is two-fold. On the one 
hand, we would like to ``validate'' our Artificial Life system by comparing
our experimental results to theoretical predictions known to describe
natural systems, such as waves of RNA strings replicating in 
Q$\beta$-replicase~\cite{BMO,MB}. On the other hand, this benchmark allows us 
to determine the diffusion coefficient and  velocity of information propagation
from relative fitness and mutation rate. Finally, we arrive at an estimate 
of the minimum system size which guarantees that the population will not, 
on average, equilibrate. 

In the next section we briefly describe {\sf sanda} and 
its main design characteristics. The third section introduces the 
reaction-diffusion equation for a discrete system and analytical results
for the wavefront velocity as a function of relative fitness and mutation
rate. 
We describe our results in the subsequent section and close with some comments 
and conclusions.

\section{The Artificial Life System ``{\sf Sanda}''}
Like {\sf avida}, {\sf sanda} works with a population of strings of code 
residing on an $M \times N$ grid with periodic boundary conditions. 
Each lattice point can hold at most one string. Each string 
consists of a sequence of instructions from a user-defined set. These
instructions, which resemble modern assembly code and can be executed 
on a virtual CPU, are designed to allow self-replication. The set of
instructions used is capable of universal computation.

Each string has its own CPU which executes its instructions 
in order. A string self-replicates by executing instructions
which cause it to allocate memory for its child, copy its own
instructions one by one into this new space, and then divide the child
from itself and place it in an adjacent grid spot. The child then
is provided with its own virtual CPU to execute its instructions.

When a string replicates, it places its child in one of the eight
adjacent grid spots, replacing any string which may have been there.
Which lattice point is chosen can be defined by the user. In our experiments,
we have
used both random selection and selection of the oldest string in the
neighbourhood. As we shall see, the selection mechanism has a significant
effect on the spread of information.

It should be noted that this birth process, and indeed all interactions
between strings, are local processes in which only strings adjacent
to each other on the grid may affect each other directly.
This is important as it both supplies the structure needed for studies
of spatial characteristics of populations of self-replicating
strings of code, and allows longer relaxation times -- making possible studies
of the equilibration processes of such systems and their nonequilibrium
behavior.

This process of self-replication is subject to mutations or errors which
may lead to offspring different from the original string and in most
cases non-viable (\ie not capable of self-replication). Of the many
possible ways to implement mutations, we have used only copy errors ---
every time a string copies an instruction there is a finite chance
that instead of faithfully copying
the instruction, it will instead write a randomly chosen one.
This chance of mutation is implemented as a mutation rate $R$ -- 
the probability of copy-error per instruction copied. A mutation rate
$R$ for a string of length $\ell$ will therefore lead to a fidelity
(probability of the copied string being identical to the original) 
$\alpha = (1-R)^\ell$.  
This then, allows us to evolve a very heterogeneous population from an
initially homogeneous one. The resulting evolution, coevolution, 
speciation etc. have been and continue to be 
studied.~\cite{TRAY2,CA1,CA3,CA2,AB1,ABH}

What decides whether one particular sequence of instructions (or genotype)
will increase or decrease in number are the rate at which it replicates,
and the
rate that it is replaced at. In our model, the latter is genotype
independent (the ``chemostat'' regime).
Accordingly, we define the former (\ie its average replication rate)
as the genotype's 
fitness.
In other words, fitness 
is equal to the inverse of the time required to 
reproduce (gestation time).

To consistently define a replication rate, it is necessary to define a 
unit of time. 
Previously, in
{\sf tierra} and {\sf avida}, time
has been defined in terms of instructions executed for the whole
population (scaled by the size of the population in the case of {\sf avida}). 
In {\sf sanda}, we define a physical time by stipulating that it takes
a certain finite time for a cell to execute an instruction.
This base 
execution time may vary for different instructions (but is kept constant in all
experiments presented here). 
The {\em actual} time a cell takes to execute a certain instruction is then
increased or decreased by changing its ``efficiency''.
Initially, each cell is assigned an efficiency near unity,
$e=(1+\eta)$, where $\eta$ represents a small stochastic component. 
In summary, the time it takes a cell to execute a 
series of instructions
depends on the number of instructions, the particular instructions
executed, and the cell's efficiency. 

Self-replication consists of
the execution of a certain series of instructions by the
cell. Thus, the fitness of the cell (and its respective genotype) is just
the rate at which this is accomplished and depends explicitly on the
cell's efficiency. 
We can assign better (or worse) 
efficiency
values to cells which contain certain instructions or which manage to 
carry out certain operations
on their CPU register values. This allows us to influence the
system's evolution so as to evolve strings which carry out allocated
tasks. A cell that manages a user-defined task can be 
assigned a better efficiency for accomplishing it. Such cells, by virtue
of their higher replication rate, would then 
have an evolutionary advantage over other
cells and force them into extinction. At the same time, the discovery 
that led to the better efficiency is propagated throughout the population
and effectively frozen into the genome. 

In addition to the introduction of a real time, {\sf sanda} differs from its
predecessors in its parallel emulation algorithm. Instead of using a
block time-slicing algorithm to simulate multiple virtual CPUs, {\sf sanda}
uses a localized queuing system which allows perfect simulation of
parallelism.

Finally, {\sf sanda} was written to run on both parallel processors and single
processor machines.  
Therefore, it is possible, using parallel computers, 
to have very large populations of strings coevolving. This permits
studies of extended spatial properties of these systems of self-replicating
strings and holds promise of allowing us to study them away from
equilibrium.

\section{Diffusion and Waves}

Information in {\sf sanda} is transported mainly by self-replication. When a
string divides into an adjacent grid site, it is also transferring the
information contained in its code (genome) to this site. We have looked
at the mode and speed of this transfer in relation to the fitness of
the genotype carrying the information, the fitness of the other
genotypes near this carrier, and the mutation rate. 

Consider what happens when one string of a new genotype appears
in an area previously populated by other genotypes. 
We will make the
assumption that the fitness of the other viable (self-replicating)
genotypes near the carrier are approximately the same. This holds for
cases where the carrier is moving into areas which are in local equilibrium.
We will use $f_c$ for the fitness of the newly introduced (carrier) genotype
and $f_b$ for the fitness of the background genotypes. If $f_c < f_b$,
obviously the new genotype will not survive nor spread.

In the following, we have studied three different cases: 
diffusion, wave propagation,
and wave propagation with mutation. 

The diffusion case represents the limit where the fitness of both 
genotypes are the same. 
It turns out that this can be modelled as a classical random walk. 
On average, if the
carrier string replicates it will be replaced before it can
replicate again. This is effectively the same as the carrier
string {\it moving} one lattice spacing in a random direction
chosen from the eight available to it. 
The random
walk is characterized by the disappearance of the mean displacement and
the linear dependence on time of the mean squared displacement:
\be
\la r\ra(t) & = & 0 \\
\la r^2\ra(t) & = & 4Dt
\ee
where $D$ is defined as the diffusion coefficient. 

For our particular choice of grid and replication rules, we find for the
diffusion coefficient of a genotype with fitness $f$,
\be
D^{(b)} = \frac38 a^2 f \label{diffb}
\ee
where $a$ is the lattice spacing. This holds for a ``biased'' selection
scheme where we select the oldest cell in the neighbourhood to be
replaced. (See below.)

If $f_c > f_b$ then we find that instead of diffusion we obtain a roughly 
circular population wave of the new genotype
spreading outward. We are interested in the speed of this
wavefront. 

Let us first treat the case without mutation.
If the radius of this wavefront is not too
small we can treat the distance from the center of the circle $r$ as a linear
coordinate. We define $\rho(r, t)$ as the mean normalized population density of
strings of the new genotype at a distance $r$ from the 
center at a time $t$ measured
from our initial seeding with the new genotype. We assume that the ages of cells
near each other have roughly the same distribution and that this distribution
is genotype independent, ensuring that the selection of cells to be replaced 
does not depend
on genotype either.

Then, we can write a flux equation (the reaction-diffusion equation) which 
determines the change in the population density $\rho(r,t)$ as a function
of time
\be
\lefteqn{\frac{\del\rho(r,t)}{\del t} = } \nonumber  \\
	& &\hspace{-0.6cm}\left[\frac38\rho(r-a,t)+\frac14\rho(r,t)
	+\frac38\rho(r+a,t)\right]\,f_c\,(1-\rho(r,t)) \nonumber \\
&-& \left[\frac38(1-\rho(r-a,t))+\frac14(1-\rho(r,t))\right. \nonumber \\
&+& \left.\frac38(1-\rho(r+a,t))\right]f_b\rho(r,t)\;.
\ee 

Since we are interested in the speed of the very front of the wave, we
can assume $\rho$ to be small. Also, from physical considerations we
assume $\rho$ is reasonably smooth. Then,
we can use a Taylor expansion for $\rho(r\pm a,t)$ and keep the lowest order 
terms to obtain
\be
\frac{\del \rho(r,t)}{\del t} = \frac38\,a^2\,f_c\,
\frac{\del^2\rho(r,t)}{\del r^2} + (f_c-f_b)\rho(r,t)\;.
\ee
This can be solved for the linear wavefront speed $v^{(b)}$ yielding~\cite{CH}
\be
v^{(b)} & = & a\,\sqrt{\frac32}\,\sqrt{f_c(f_c-f_b)}  \\
 & = & 2\sqrt{D^{(b)}_c(f_c-f_b)}
\label{vb0}
\ee
where $D^{(b)}_c$ is the diffusion coefficient of the carrier genotype
when using a biased (by age) selection scheme.

To study the case of wave-propagation with mutation we shall make the 
assumption that all mutations
are fatal. We can then calculate a steady state density of non-viable
cells $\delta$,
\be
\delta = 1-\alpha^{1/8}
\ee
where the fidelity $\alpha$ is the probability that a child will have the 
same genotype as 
its parent ({\it i.e.}, not be mutated). As mentioned earlier, the
fidelity is related to the 
mutation rate $R$ by
\be
\alpha = (1-R)^\ell 
\ee
where $\ell$ is the length of the particular string.
Modifying our previous flux equation to take into account these new factors
and repeating our previous analysis gives us
\be
v^{(b)} = 2\,\alpha\sqrt{D^{(b)}_c(f_c-\alpha^{1/8}f_b)}\;.
\label{vbr}
\ee

Let us now consider the effects of different selection schemes for
choosing cells to be replaced.
The relations we derive above hold true for the case in which we replace
the oldest cell in the 8-cell neighbourhood when replicating
(``age-based'' selection). Another method 
of choosing a cell for replacement is to choose a random neighbouring cell
regardless of age.
This scheme, which we term ``random selection'' as opposed to the biased 
selection treated above, effectively halves the replication rate of
all cells. It follows that the diffusion coefficient is also halved,
\be
D^{(r)} & = & \frac{3}{16} a^2 f \label{diffr} \\
& = & \frac12 D^{(b)}
\ee
and for the velocity of the wavefront (with no mutation) we find
\be
v^{(r)} = 2\sqrt{D^{(r)}_c\frac{(f_c-f_b)}2}\;.
\label{vrr}
\ee
In Fig.1, we show a histogram of the number of offspring
that a cell obtains before being replaced by a neighbour's offspring,
for the biased selection case (left panel) and the random 
case (right panel). As expected from general arguments, half of the cells 
in the random selection scenario are replaced before having had a chance to 
produce their first offspring (resulting in a reduced diffusion coefficient), 
while biased selection ensures that most cells
have exactly one child.
\begin{figure*}
	\centerline{\psfig{figure=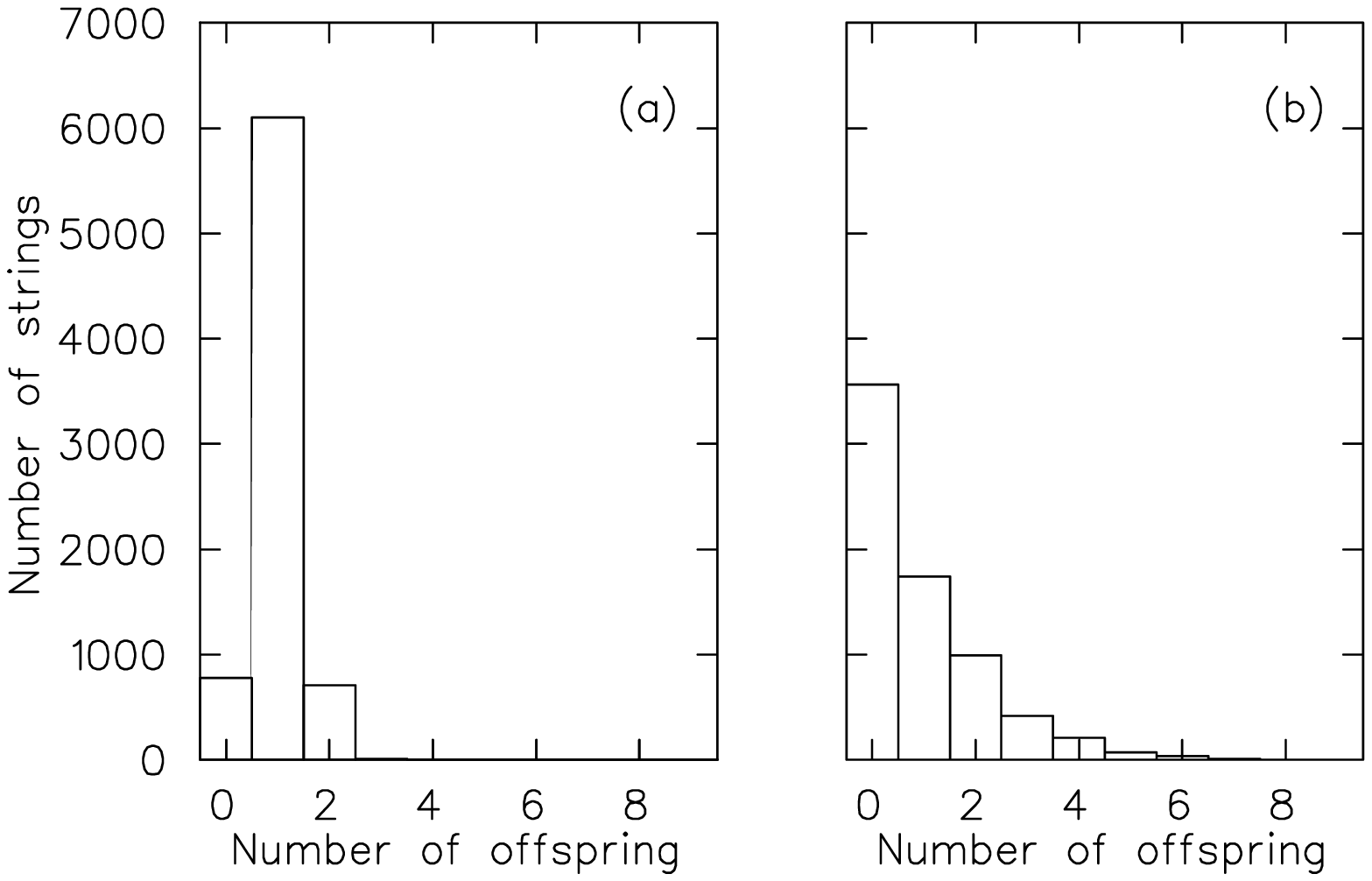,width=4.0in,angle=90}}
	\begin{quotation} 
		\noindent FIG. 1\ \ \ Distribution of number of strings generating different
		numbers of offspring, for the biased selection case [panel (a)] and
		the random selection scenario (b).
	\end{quotation}
\end{figure*} 

\section{Results}

We carry out our experiments by first populating the grid with 
a single (background) genotype of fitness $f_b$. Then, a single
string of the carrier genotype with fitness $f_c$ is
placed onto a point of the grid at time $t=0$. We then observe
the position and speed of the wavefronts formed, the mean squared
displacement of the population of carrier genotypes, and various other
parameters as a function of time.

With $f_b$ kept constant\footnote{The gestation time was approximately 
330,000,
where the base execution time for each instruction was (arbitrarily) set to 
1000: $f_b = \frac{1}{330000}$},
we have varied
${f_b}/{f_c}$ from 0.1 to 1.0 in increments of 0.1. Also, the mutation
rate $R$ was varied from 0 to $14\times 10^{-3}$ mutations per instruction,
in increments of $1\times 10^{-3}$.

A comparison of the theoretical {\it vs.} measured mean square displacement
as a function of time
for a genotype with no fitness advantage compared to its neighbours
(${f_b}/{f_c} = 1$)
is shown in Fig.2. The data were obtained from approximately 1500 runs.
The solid lines represent the (smoothed) averages of our measurements
(for biased and random selection schemes), while
the dashed lines are the theoretical predictions obtained from the diffusion
coefficients (\ref{diffb}) and (\ref{diffr}) respectively.
The slopes of the measured and predicted lines
agree very well confirming the validity of our random walk model
and the diffusion coefficient predicted by it (without any free parameters). 
The slight discrepancy
between the experimental curves and the predicted ones at small times is due
to a finite-size effect that can be traced back to the coarseness of the grid. 
\begin{figure*}
\centerline{\psfig{figure=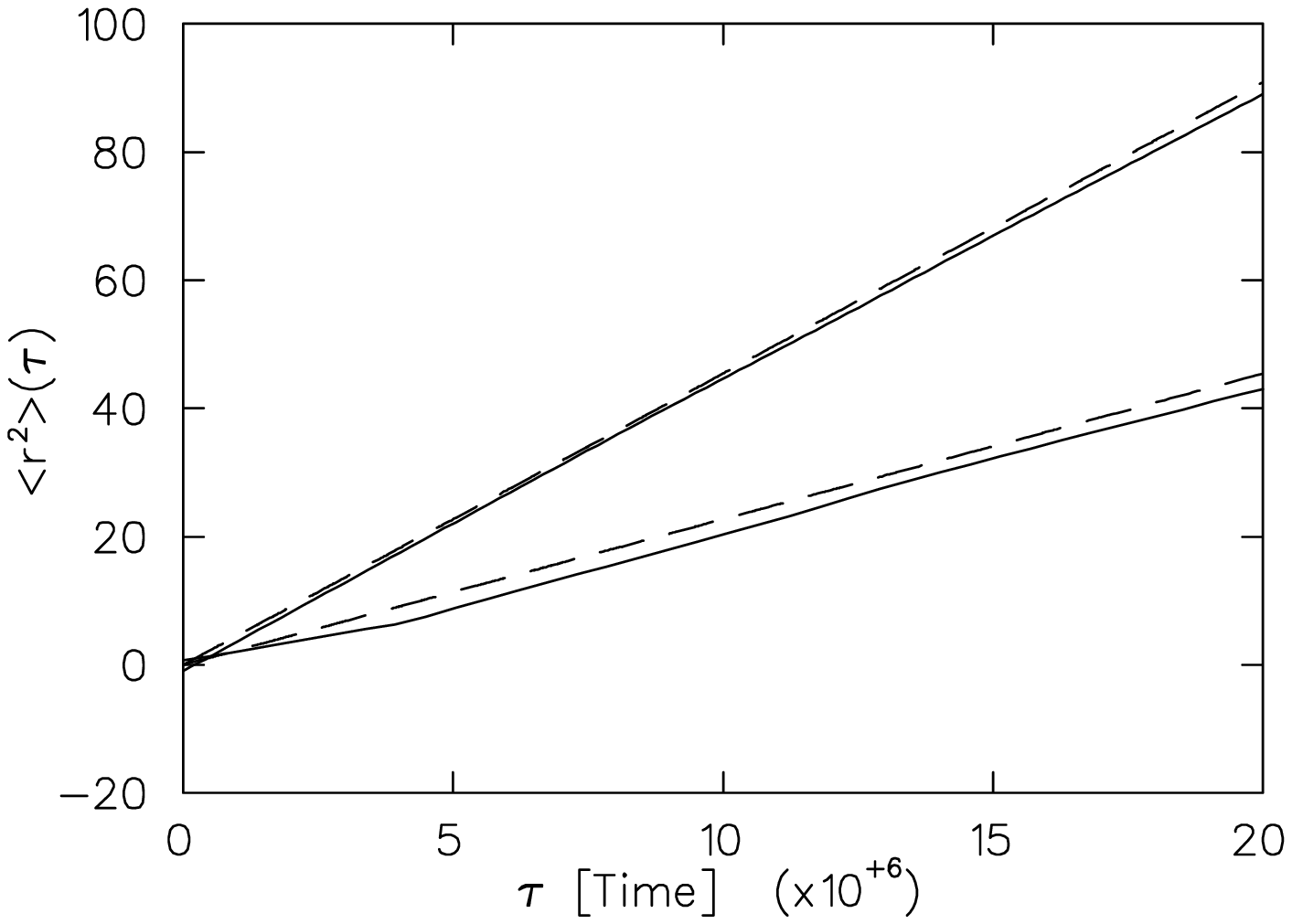,width=4.0in,angle=90}}
\begin{quotation}
\noindent FIG. 2\ \ \ Mean squared displacement of genome as a function of 
time due to diffusion. Solid lines represent experimental results obtained
from 1500 independent runs. Dashed lines are theoretical predictions. The upper
curves are obtained with the biased selection scheme while the lower curves 
result from the random selection scenario.
\end{quotation}
\end{figure*}  

Fig.3 shows the measured values of the wavefront speed for cases
where $f_c > f_b$ and without mutation, with the corresponding
predictions. Again, the higher curve is for biased and
the lower for random selection. Note that the wavefront speed gain from an
increase in fitness ratio is much better than linear. Note also that 
all predictions are again free of {\em any} adjustable parameters. 
\begin{figure*}
\centerline{\psfig{figure=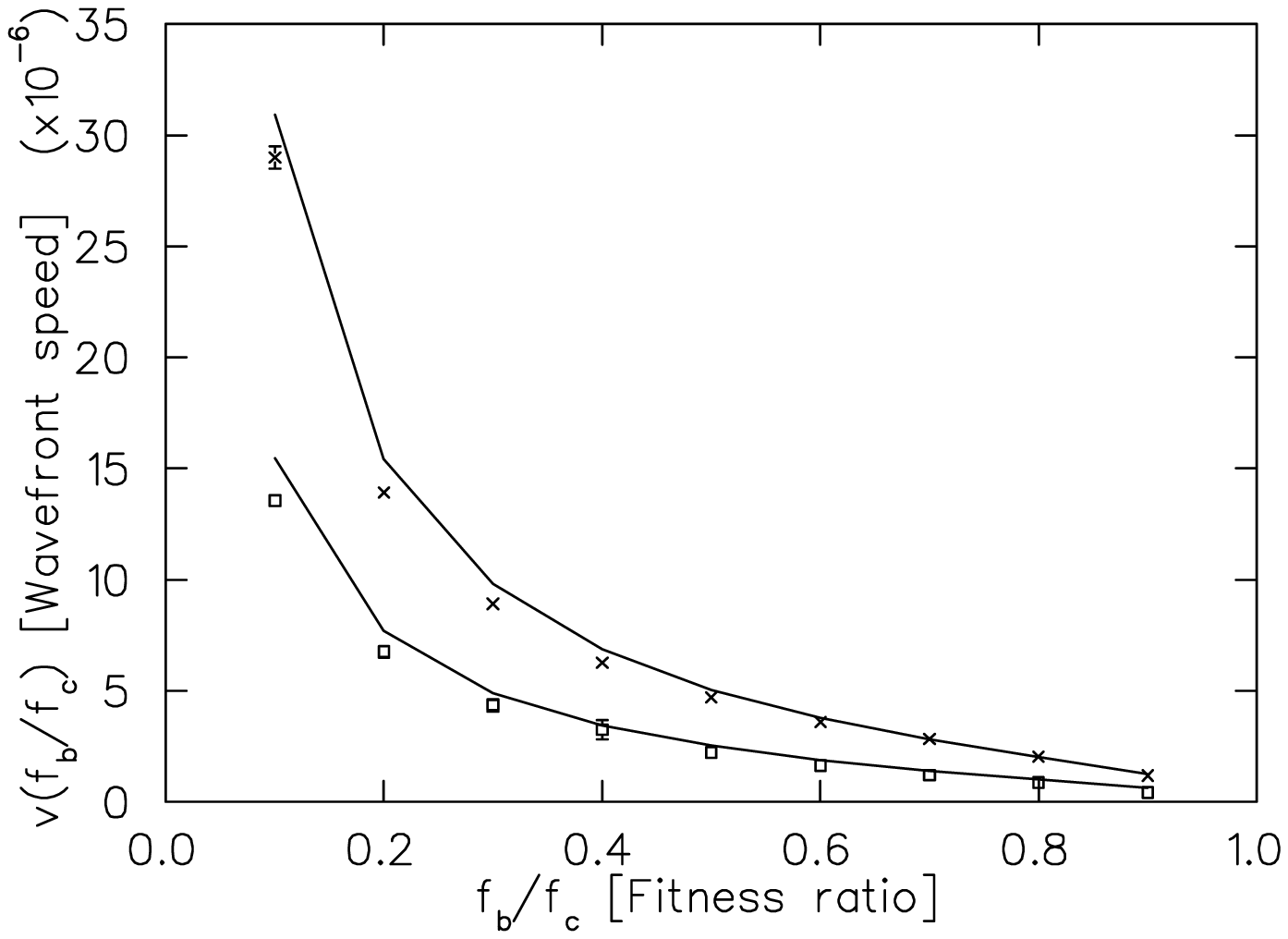,width=4.0in,angle=90}}
\begin{quotation}
\noindent FIG. 3\ \ \ 
Wavefront speed of a genotype with fitness $f_c$ propagating through a 
background of genotypes with fitness $f_b$, averaged over four runs
for each data point. Upper curve: biased selection, lower curve: random 
selection. Solid lines are predictions of Eqs. (\ref{vb0}) and (\ref{vrr}). 
\end{quotation}
\end{figure*}  

The dependence of this curve on the mutation rate is shown in
Fig.4. Increasing the mutation rate tends to push the speed of the wave
down. It should be noted, however, that because we have only used
copy mutations there is no absolute cutoff point or error threshold $\alpha_c$
where all genotypes cease to be viable, with $\alpha_c>0$. Rather, 
genotypes can spread until $\alpha$ is very close to the limit $\alpha_c = 0$.
\begin{figure*}
\centerline{\psfig{figure=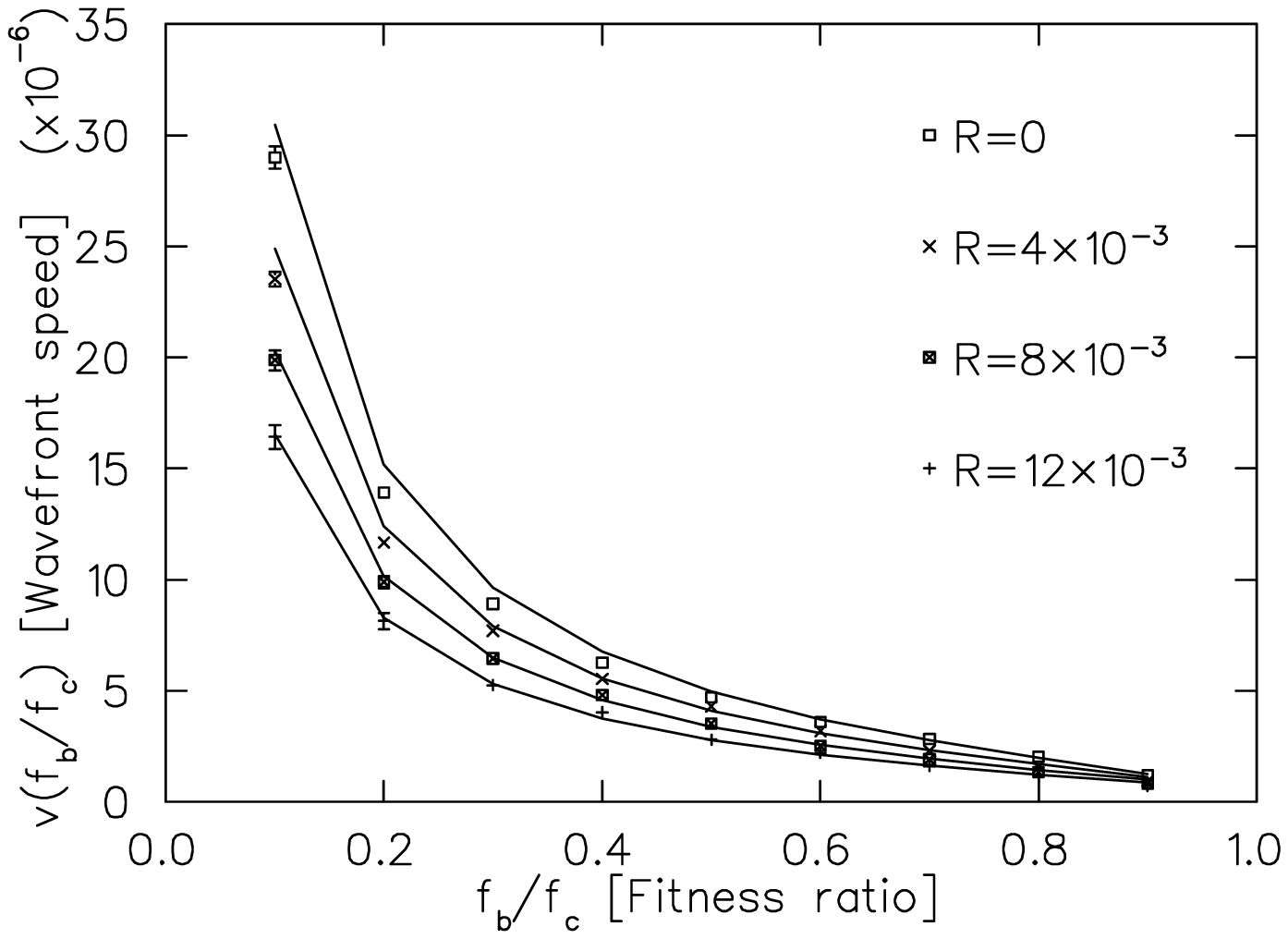,width=4.0in,angle=90}}
\begin{quotation}
\noindent FIG. 4\ \ \ 
Measured wavefront speeds versus fitness ratio for selected mutation rates $R$
(symbols) are plotted with the theoretical predictions from Eq.~(\ref{vbr})
(for the biased selection scheme only). 
\end{quotation}
\end{figure*}  

Finally, we plot the dependence of the wavefront speed on the mutation
rate for a fixed value of the fitness ratio (${f_b}/{f_c} = 0.6$)
in Fig.5. Data were obtained from an average of four runs per point in 
the biased selection scheme. Again, the prediction based on the 
reaction-diffusion equation with mutation agrees well (within error bars) 
with our measurements.
\begin{figure*}
\centerline{\psfig{figure=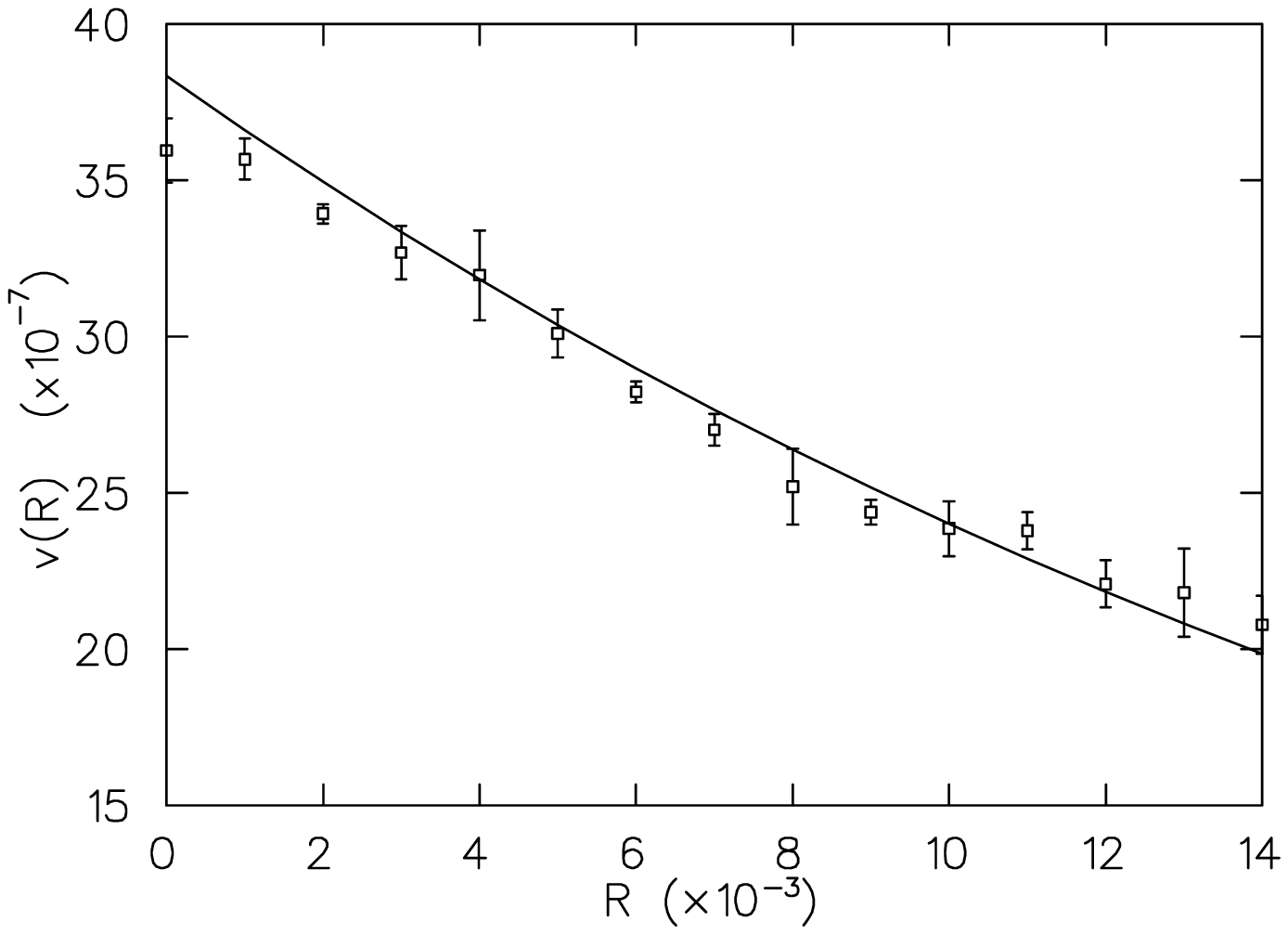,width=4.0in,angle=90}}
\begin{quotation}
\noindent FIG. 5\ \ \ 
Wavefront speed of a genotype (biased selection) with relative 
fitness $f_b/f_c = 0.6$ as a 
function of mutation rate (symbols). Solid line is prediction of 
Eq.~(\ref{vbr}). 
\end{quotation}
\end{figure*}  

\section{Discussion and Conclusions}

Information propagation via replication into physically adjacent sites
can be succinctly described by a reaction-diffusion equation.
Such a description has been used in the description of {\em in-vitro} 
evolution of RNA replicating in Q$\beta$-replicase~\cite{BMO,MB}, as well 
as the replication of viruses in a host environment~\cite{YM}. The same 
equation is used to describe the wave behavior of different strains
of {\em E. Coli} bacteria propagating in a petri dish~\cite{AGL}, even 
though the means of propagation in this case is motility rather than 
replication. 

We have constructed an artificial living system ({\sf sanda}) 
based on the {\sf avida} design which allows the investigation of large
populations of self-replicating strings of code, and the observation of 
non-equilibrium effects. The propagation of information was observed
for a broad spectrum of relative fitness, ranging from the diffusion
regime where the fitnesses are the same through regimes where the
difference in fitness led to sharply defined wavefronts propagating
at constant speed. The dynamics of information propagation led to
the determination of a crucial time scale of the system which represents 
the average time for the system to return to an equilibrium state 
after a perturbation. This relaxation time
depends primarily on the size of the system, and the speed of information
propagation within it. Equilibration can only be achieved if
the mean time between (non-lethal) mutations is larger than the
mean relaxation time. Thus, a {\em sufficiently} 
large system will never be in equilibrium. Rather, it is inexorably 
driven far from equilibrium by persistent mutation pressure.

For artificial living systems such as the one we have investigated, it is 
possible to formulate an approximate condition which ensures that it will 
(on average) never equilibrate, but rather consist of regions of local
equilibrium that never come into informational contact. From the timescales 
mentioned above, we determine that the number of cells $N$ in such a system 
must exceed a critical value:
\be
N > \left(\frac{2\,v(f)}{R_\star\,a}\right)^{2/3}\;,
\ee
where $R_\star$ is the rate of {\em non-lethal mutations}, $v(f)$ the 
velocity of information waves, and $a$ the lattice spacing (assuming
a mean time between non-lethal mutations $t_\star\approx(N\,R_\star)^{-1}$).

Beyond the obvious advantages of a non-equilibrium regime for genomic 
diversity and the origin of species, such circumstances offer the
fascinating opportunity to investigate the possibility of non-equilibrium
pattern
formation in (artificial) living systems. However, the most interesting
avenue of investigation opened up by such artificial systems is that of
the study of the fundamental characteristics of life itself. Since it is 
widely believed that many of
the processes that define life, including evolution, occur
in a state which is far from equilibrium, to study such processes it is
necessary to have systems which exhibit the properties of life we
are interested in and that can be quantitatively studied in a 
rigorous manner in this regime. The availability of artificial living 
systems as experimental testbeds that can be scaled up to arbitrary
population sizes on massively parallel computers is a step in this direction.

\section*{Acknowledgements}
J.C. would like to thank Mike Cross for 
continued support, 
and Roy Williams,
Than Phung and the Center for Advanced Computing Research at Caltech
for their help.
C.A. was supported 
in part by
NSF grants PHY94-12818 and PHY94-20470. 
This research was performed in part using the CSCC parallel computer system
operated by Caltech on behalf of the Concurrent Supercomputing Consortium.
Access to this facility was provided by Caltech.
\bibliographystyle{unsrt}

\begin{thebibliography}{99}
\bibitem{FIS} R.A. Fisher, Ann. Eugen {\bf 7} (1937)355.
\bibitem{DOB} Th. Dobzhansky and S. Wright, Genetics {\bf 28} (1943)304.
\bibitem{BMO} G.J. Bauer, J.S. McCaskill, and H. Otten, Proc. Natl. Acad. Sci.
USA {\bf 86} (1989)7937.
\bibitem{MB} J.S. McCaskill and G.J. Bauer, Proc. Natl. Acad. Sci.
USA {\bf 90} (1993)4191.
\bibitem{RAY1} T. S. Ray, in {\it Artificial Life II:} Proceedings of 
an Interdisciplinary Workshop on the Synthesis and Simulation of Living
Systems, Santa Fe Institute Studies in the Sciences of Complexity, Proc. 
Vol.~10, edited by C. G. Langton et al., Addison-Wesley, 
Reading, MA, p. 371 (1992).
\bibitem{TRAY2} T. S. Ray, Physica {\bf D 75} (1994)239; Artificial Life {\bf 1}
(1994)195.
\bibitem{CA1} C. Adami, Physica {\bf D 80} (1995)154.   
\bibitem{CA3} C. Adami, Artificial Life {\bf 1} (1994)429.
\bibitem{CA2} C. Adami, Phys. Lett. {\bf A 203} (1995)23.
\bibitem{AB1} C. Adami and C.T. Brown, In R.A. Brook and P. Maes (Eds.), 
{\em Artificial Life IV}: Proceedings of the Fourth International Workshop
on the Synthesis and Simulation of Living Systems, p. 377. 
MIT Press, Cambridge, MA (1994).  
\bibitem{ABH} C. Adami, C.T. Brown, and M. Haggerty, Proc. of 3rd
Europ. Conf. on Artificial Life, June 4-6, 1995, Granada, Spain, 
Lecture Notes in Computer Science p.503, Springer Verlag (1995). 
\bibitem{CH} M. C. Cross and P.C. Hohenberg, Rev. Mod. Phys. {\bf 65} (1993)851.
\bibitem{YM} J. Yin and J.S. McCaskill, Biophys. J. {\bf 61} (1992)1540.
\bibitem{AGL} K. Agladze et al., Proc. Roy. Soc. Lond. B, {\bf 253} (1993)131. 

\end{thebibliography}

\newpage
\end{document}